# Porous Structured Au Colloids: Insights in to Morphology, Optical and Antimicrobial Activity

G. Nirmala Devi[a], R. N. Viswanath[a,*], C. Lakshmanan[b], G. Suresh[a], R. Rajaraman[b]

[a]Centre for Nanotechnology Research, Department of Humanities and Science
Aarupadai Veedu Institute of Technology, Vinayaka Mission's Research Foundation, Paiyanoor, Chennai - 603 104, TamilNadu, India

[b]Materials Science Group, Indira Gandhi Centre for Atomic Research, HBNI, Kalpakkam 603 102, Tamil Nadu, India

**Abstract:** Porous structured Au colloids have been prepared from bulk nanoporous Au by means of an element dissociation method. The microscopic techniques (scanning electron microscope and atomic force microscope), UV-Vis spectroscopy and zone diffusion method have been employed to study their morphology, optical property and anti-microbial activity against Gram-positive and Gram-negative bacterial strains, respectively. It is shown that the porous structured Au colloidal suspension exhibit excellent optical and antimicrobial properties. The noticeable features present in the optical studies are two Plasmon resonance peaks at 477 and 546 nm with overlapping of these peaks at a wavelength of 520 nm. The morphology studies by SEM and AFM indicate that the Au colloids are rod shaped with an assembly of skeletal pore and ligament structure. The analysis of the antimicrobial activity testing reveals that the porous structured Au colloids have greater inhibitory effect against the tested Gram-positive and Gram-negative bacterial strains.

**Key words:** Au colloids; porous structure; zone agar diffusion; optical resonance; Antimicrobial activity

[*] Corresponding Author

# 1. Introduction

The R & D activity on nanoporous metals such as nanoporous Au for biomedical application is progressing recently due to their special properties in different aspects [1, 2, 3, 4]. However, there are a limited data available in the literature discussing about the effectiveness of its antimicrobial activity. This is mainly because of the experimental difficulties in making highly dispersed porous structured Au colloids and conducting experiments for antimicrobial susceptibility testing. The interest of porous structured Au colloids for antimicrobial research is also increasing due to the fact that materials which are currently used in clinical practice for disinfection have disadvantages found in a number of cases, including toxicity to the human body over a period of time, giving trouble persistently by weak antimicrobial activities and difficulty in functioning in a dynamic environment [5].

Miyasawa et al. reported the antimicrobial activity of bulk nanoporous Au and demonstrated that nanoporous Au has the ability to disturb bacterial growth simply through indirect interactions between the walls of bacteria and the surface of nanoporous Au without diffusing through the Bacterial cell walls [6]. Because of their bi-continuous microstructure of open porosity with interconnected metal ligaments, owing to the increased high surface area, ability to tailor the charge injection/withdraw capability, ability to modify the structure size dependent properties and compatibility to bio-fluids, nanoporous Au has received special attention in health care technology [7, 8, 9, 10]. Even though antimicrobial activity of bulk nanoporous Au has been reported, the mechanism behind their microbial activity is not hitherto well understood.

One of the current research interests lie mainly on developing porous structured Au colloidal antimicrobial agents with long term preventing anti bacterial features. The present article reports the morphology, optical property and antimicrobial activity of porous structured Au colloids against various gram positive and gram negative bacterial strains.

## 2. Materials and Methods

The Au dissociation [11] experiment was performed in a three-electrode electrochemical glass cell consisting of two identical freshly prepared nanoporous Au discs as working and auxiliary electrodes, Ag/AgCl reference electrode and 0.1 M $HClO_4$ electrolyte. Well characterised hierarchical structured nanoporous Au samples prepared by electrochemical dealloying [1, 2, 4] of binary $Ag_{70}Au_{30}$ were used for the Au dissociation experiments. Since the Au dissociation occurs above the oxide formation potentials [11], the potential sweep was performed successively upon cycling the potential intervals $1.1 < E < 1.7$ V versus Ag/AgCl at a scan rate of 1 mV/s. for 200 cycles. The Au colloids obtained from the cyclic voltammetry experiment were cleaned by following standard procedures and used for further studies. The optical property of the Au colloidal suspension in isopropyl alcohol was analyzed using UV-Visible spectrometer (UV-Vis 2600, Shimadzu). Since the morphology of Au colloids could have an impact on its antimicrobial activity, a droplet of the Au colloidal suspension was spread on silicon single crystal surface, dried and investigated through AFM (Park XE7) and scanning electron microscope (Carl Zeiss Supra 55).

The antimicrobial activity testing of porous structured Au colloidal solution was performed against different Gram-positive strains such as Bacillus subtilis, Micrococcus luteus, Staphylococcus aureus and Gram-negative strains such as Escherichia coli, Proteus vulgaris, Shigella flexneri. The Agar well diffusion method [12] which is a standard method for quantifying the ability to inhibit bacterial growth was used for the antimicrobial activity testing. The aseptic chamber which consists of a wooden box of dimensions 1.3 m x 1.6 m x 0.6 m with a door closure was cleaned with 70% ethanol and irradiated with short wave length UV light source. The Nutrient broth agar diffusion medium [13] is used for cultivation of microorganisms in accordance with a procedure using peptone (5 g), yeast extract (3 g), sodium chloride (5 g), ultra pure water (18.2 MΩ) ( 1 L), agar (20 g). Depending upon the availability of strains, the medium was calculated and suspended in 200 mL of ultra-pure water in a 500 mL conical flask, stirred, boiled to dissolve and then autoclaved in 15 lbs at 121ºC for 15 min [14]. The hot medium was poured in sterile Petri plates and kept under aseptic laminar air flow chamber for 15 min for solidification.

The solidified nutrient agar in the Petri plates were inoculated by dispensing the inoculum using sterilized cotton swabs and spread evenly onto the solidified agar medium. Standard wells were created in each plate with the help of a sterile well-borer of 8 mm diameter. The Au colloids in various concentrations were then poured into each well. All the plates with extract loaded wells were incubated at 37ºC for 24 h and the antibacterial activity was assessed by measuring the diameter of the inhibition zone formed around the well. Tetracycline (50 µg) was used as positive control.

## 3. Results and Discussion

Figure 1 a) shows the cyclic voltammetry spectra recorded in the cell potential intervals $1.1 < E < 1.7$ $V_{Ag/AgCl}$ during the dissociation of Au colloids from the bulk nanoporous Au. Three selected scans from the 200 successive scans (scan nos.10, 70 200) are depicted in the figure. It is seen that there is an upsurge in the current flow above 1.35 $V_{Ag/AgCl}$, which is in the oxygen evolution region. Cherevko et al. through their combined micro-electrochemical and inductively plasma mass spectrometry experiments demonstrated that at higher positive potentials (> 1.6 $V_{RHE}$, oxygen evolution reaction takes place along with concomitant Au loss by the electrochemical reaction Au $\rightarrow$ $Au^{3+}$ + 3 $e^-$ [11]. This indicates that the main cause for the pronounced rise of the current wave above 1.35 $V_{Ag/AgCl}$ in the anodic sweep is the dominance of Au dissociation (cf.: Fig. 1a)). A photograph depicts in Fig. 1 b) illustrates the resultant

`4

porous structured Au colloidal suspension in isopropyl alcohol in four different concentrations 125, 200, 250 and 375 µg/mL.

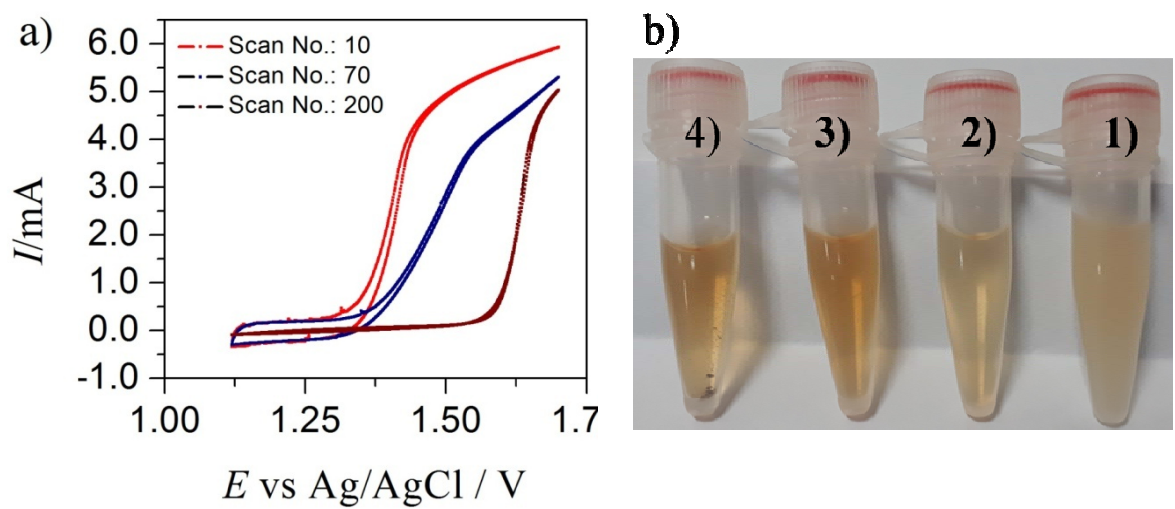

Fig: 1: a) Cyclic voltammetry spectra recorded in the cell potential intervals $1.1 < E < 1.7$ $V_{Ag/AgCl}$ during the dissociation of Au colloids from the bulk nanoporous Au. Potential sweep was performed for 200 cycles. Three selected cyclic voltammograms from the two hundred successive scans (scan nos.: 10, 70 and 200) are depicted. b) A photo showing Au colloids in four different concentrations 125, 200, 250, 375 µg/mL.

The photo in Fig. 2 shows an example of an antimicrobial sensitivity testing of porous structured Au colloidal suspension in various concentrations against Gram-positive strain (Staphylococcus aureus). Similar results were obtained on other tested Gram-positive strains Micrococcus luteus, Bacillus subtilis, and Gram-negative strains Escherichia coli, Shigella flexneri, Proteus vulgaris. It is seen from the zone of inhibition in cultures in Petri plates that the dispersed porous structured Au has antimicrobial effect against the Gram positive and Gram negative strains. Table 1 summarizes the antimicrobial efficacy of porous structured Au suspension. The quantitatively measured diameter of zone of inhibition data reveals that the antibacterial sensitivity of the porous structured gold suspension and its potency are found to vary depending up on the nature of microbial species. A Maximum inhibition zone of 18 mm for Gram-positive strain (Staphylococcus aureus) and 16 mm for Gram-negative strain (Shigella flexneri) were obtained at 375 µg/mL concentration.

To gain further insight in to the morphology of Au colloidal particles, the SEM and AFM micrographs depicted in Fig. 3 are discussed. Figures 3 a), b)) show the SEM micrographs of dried Au colloids in two distinct magnifications. It is seen from the SEM image at low magnification (Fig. 3 a) that the Au colloids are mostly in the form of cylindrical rods. The high magnified view of the SEM micrograph (Fig. 3 b)) revealed that the rod shaped Au colloids are in the form of a bi-continuous pore and ligament interconnected structure, which is a typical microstructure reported for nanoporus Au. Here, it is worth reporting that at higher magnification, the image contrast of porous structured Au colloids is poor when compared to that of starting nanoporous Au discs (Fig. 3 c)) although identical imaging parameters were chosen. This substantial discrepancy in the image contrast might either be due to surface Z-contrast or evolution of fine structures on the porous structured Au colloids. The typical image of Au colloids recorded during the scanning of the AFM tip across the sample in an observation area of 10 x 10 µm$^2$ is shown in Figs. 3 d). The aspects ratio of the ligaments (AR$_l$) estimated from the ratio of mean ligament length $l$ to its mean diameter $d$ in the rod specimen yields 1.3.

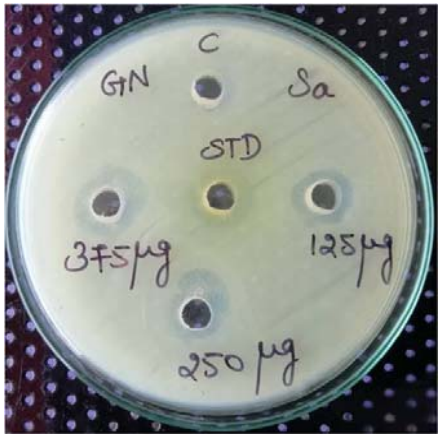

Fig. 2: Typical antimicrobial efficacy of porous structured Au colloidal suspension in various concentrations 125, 250 and 375 µg/mL against a Gram-positive strain (Staphylococcus aureus). It is noting that the result corresponding to 200 µg/mL is not displayed.

`6

TABLE 1: Antimicrobial activity (zone of inhibition) of porous structured Au colloidal suspension in various concentrations against six bacterial strains

| Bacterial Strains | Zone of inhibition (mm) | | | | |
|---|---|---|---|---|---|
| | 125 µg/mL | 200 µg/mL | 250 µg/mL | 375 µg/mL | Standard (Tetracycline) |
| **Micrococcus luteus** | 11 | 11.75 | 12.5 | 14 | 18 |
| **Bacillus subtilis** | 10 | 10.8 | 11.5 | 13 | 14 |
| **Staphylococcus aureus** | 13.5 | 14.5 | 16 | 18 | 12 |
| **Escherichia coli** | 11.4 | 12.4 | 13.2 | 14.5 | 14 |
| **Shigella flexneri** | 12 | 12.8 | 14 | 16 | 22 |
| **Proteus vulgaris** | 10.5 | 11.25 | 12 | 13.5 | 18 |

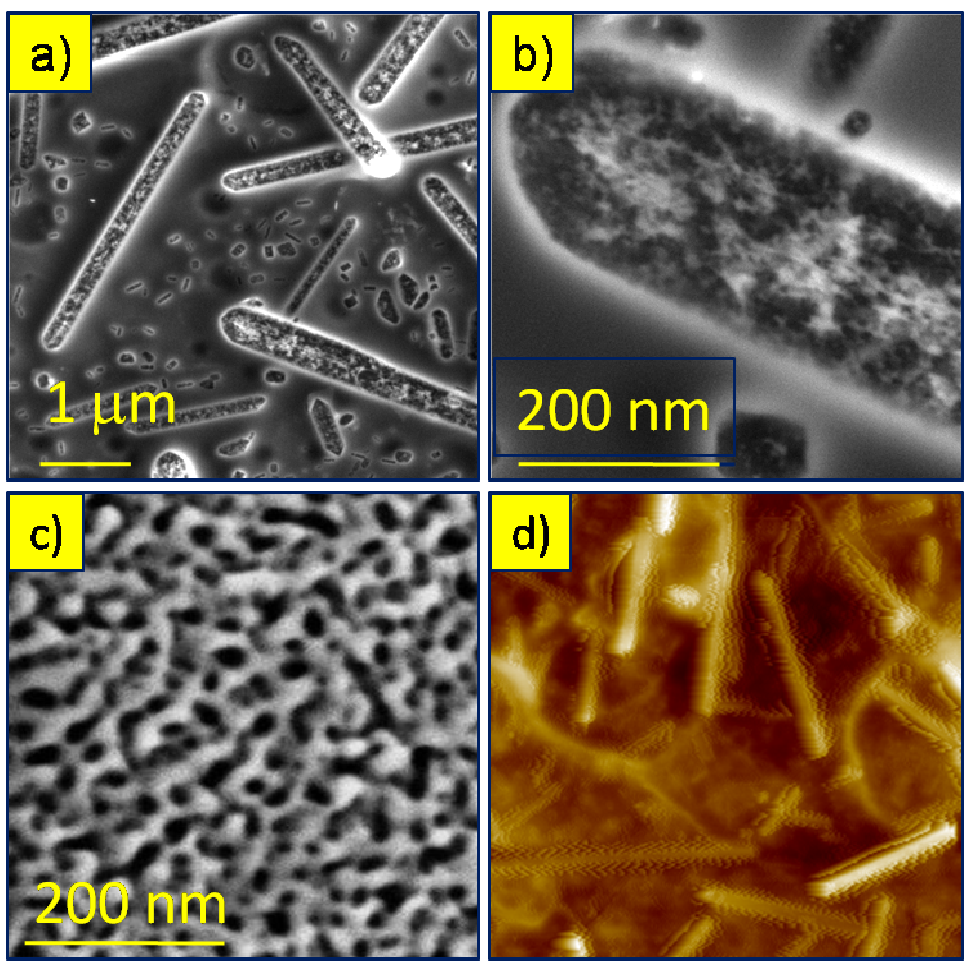

Fig. 3: SEM and AFM results of Au colloidal particles. a), b) Low and high magnified SEM images of Au colloids. c) Typical SEM image of a nanoporous Au sample used for the Au dissociation experiment.. d) Typical AFM image of Au colloidal particles in an observation area 10 x 10 μm$^2$

Figure 4 a) shows the typical optical spectrum of porous structured Au colloidal suspension recorded from 400 to 800 nm. The noticeable features present in the spectrum are two Plasmon resonance peaks centred at 477 and 546 nm. It is seen that these peaks overlap at 520 nm. Since the optical function of nm scaled objects depend highly on their size and shape, we look at the Plasmon resonance results of present and the results of earlier measurements on nanoporous Au. Jalas et al. through their observation of two distinguished spectral peaks with their overlaps in the optical measurements suggested that the sample nanoporous Au behaves as a diluted Au system [15]. Their study explores that for longer wavelengths the light is mostly transmitted axially through Au networks and for shorter wavelengths optical losses of Au becomes dominant. Since Jalas et al. predicted that the intrinsic hierarchical nanostructures influence to the optical properties in nanoporous Au, the aspect ratio value of 1.3 determined from the present microscopy studies has been compared with the models. The matching aspect ratio R value obtained from the linear relation $\lambda = (53.7*R - 42.29) \varepsilon_m + 495.14$ for the longer wavelength 546 nm in the absorption band yields 1.32 [16]. For a purpose of a qualitative comparison, we have verified the aspect ratio value obtained from the present study with one of our earlier results where ball-stick model [17] enables to determine the aspects ratio value 1.33 from the geometrical surface area value of nanoporous Au. Based on the results discussed above, we are led to conclude that the origin of the two Plasmon resonance peaks obtained in the present study are largely due to the presence of nm scaled hierarchical structures in the rod shaped Au colloids.

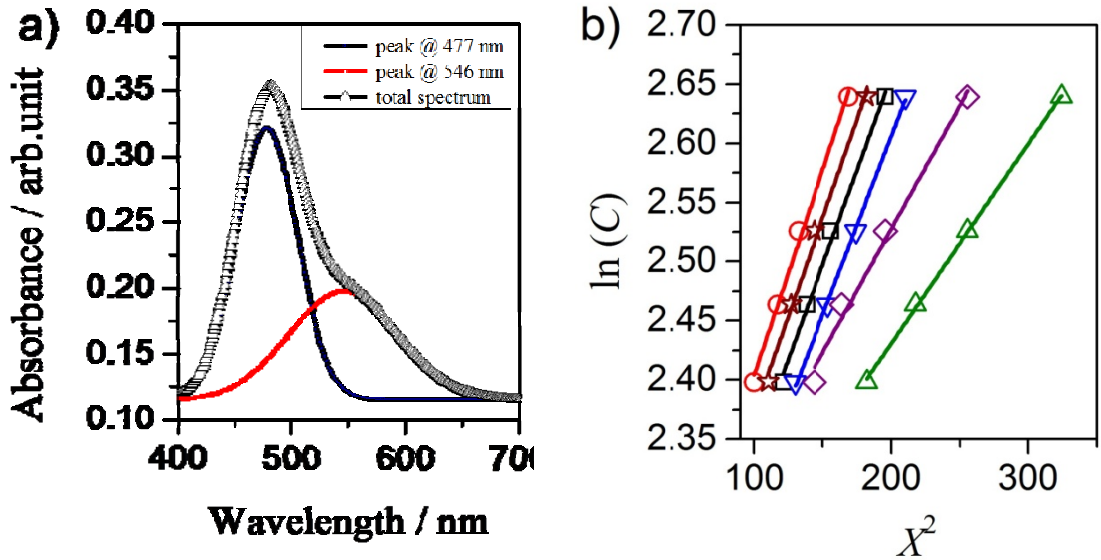

Fig. 4 a) UV-Vis optical absorption spectrum of porous structured Au colloidal suspension recorded in the range 400 – 700 nm. b) The squared values of inhibition zone radii $X^2$ is plotted for six microbial strains against logarithm of the concentration of porous structured Au colloids $ln(C)$ together with their linear fit.

It is emphasized that the motivation behind the synthesis of porous structured Au colloidal medium is to find whether it can be exploited for improved antimicrobial efficacy. To envisage this problem, the antimicrobial results summarized in Table 1 are analysed. Figure 4 b) displays a set of linear plots of concentration $ln(C)$ of porous structured Au colloids versus squared value of inhibition zones $X^2$. It is seen that these linear plots fit well with the empirical function $ln(C) = ln(MIC) + X^2/4Dt$ [18] [18]. The diffusion constant D value, which is independent of the concentration of antimicrobials [19] is $9.1 \times 10^{-6}$, $8.3 \times 10^{-6}$, $1.7 \times 10^{-5}$, $9.6 \times 10^{-6}$, $1.4 \times 10^{-5}$, $8.7 \times 10^{-6}$ cm$^2$/s for Micrococcus luteus, Bacillus subtilis, Staphylococcus aureus, Escherichia coli, Shigella flexneri, Proteus vulgaris, respectively. The minimum inhibitory concentration MIC value $2 \pm 0.1$ μg/$l$ obtained from the fit, which is indicative of the lowest concentration of porous structured Au antimicrobial that inhibits the visible growth of microorganisms, does not change appreciably with the types of microorganisms. Thus, the antimicrobial activity testing results as discussed above indicate that the rod shaped porous structured Au colloids have the ability to kill infected microbial cells (cf. Figs. 2), 6) and Table 1)).

Now we try to understand the underlying mechanisms that involved in the dysfunction of the bacterial activity in the presence of the suspended porous structured Au colloids. According to the cyclic voltammetry results summarised in Fig. 1 a), the Au dissociation occurs from the bulk nanoporous Au at higher positive potentials (> 1.35 $V_{Ag/AgCl}$) which is much beyond the oxide formation potential of Au [11]. Hence, it is considered that the Au colloids that are dissociated from the nanoporous bulk Au carry a significant amount of positive surface charge [11, 17]. These frozen positive charges in the dispersed Au colloids disturb the cell membrane by the mode of electrostatic interaction between $Au^{3+}$ ions and negatively charged cell membrane. As the interaction of bacterial cell with positively charged Au surfaces become more effective, these charged species synergistically exerts antimicrobial effect by altering membrane stability, initiate cell damage and kill the infected bacteria.

## 4. Conclusion:

Highly dispersed porous structured Au solution has been successfully synthesized by an electrochemical dissociation method and its morphology, optical properties and antimicrobial activity were studied. The origin of two Plasmon resonance peaks observed at 477 and 546 nm and their overlap at 520 nm might be due to the presence of nanometer scaled hierarchical structures in the dispersed Au rods. The antimicrobial activity testing results reported in this study revealed that the porous structured Au dispersed in aqueous medium enables to target physical structures through electrostatic interactions and have great potential to kill different kinds of bacterial strains. Though the use of porous structured Au dispersoid has demonstrated effective antimicrobial properties, understanding the exact mechanism involved for effective controlling the bacterial growth is still a topic of current debate.

**Acknowledgments:** The authors acknowledge gratefully the Vinayaka Missions Research Foundation, Chennai 603 104 for the research support and encouragement. The research and financial support from IGCAR and UGC-DAE CSR are gratefully acknowledged.

**Graphical Abstract:**

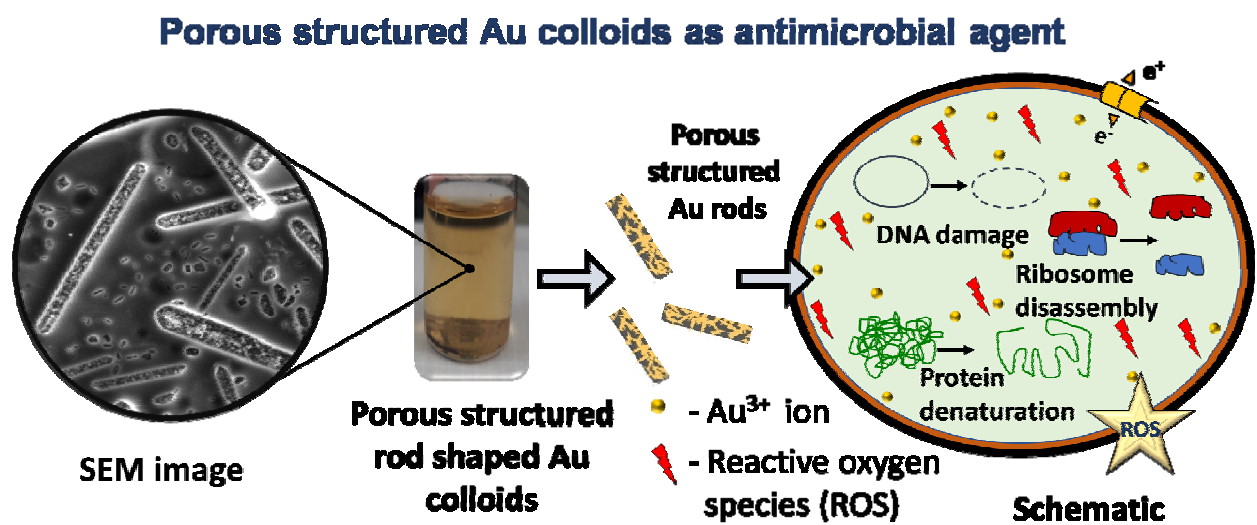

**Highlights:**

- Porous structured Au colloids in aqueous medium have been prepared by an element dissociation method.

- The origin of two distinct Plasmon resonance peaks appeared in dispersed porous structured Au colloids are largely due to the presence of nm scaled Au hierarchical structures.

- Porous structured Au colloids in aqueous medium have greater inhibitory effect against different kinds of Gram-positive and Gram-negative bacterial pathogens.

- The results summarized would make it possible to manufacture porous structured Au colloidal antimicrobial agent.